
\magnification=1200
\tolerance=10000
\hsize 14.5truecm
\hoffset 1.25truecm
\font\cub=cmbx12
\baselineskip=22truept
\parindent=1.truecm
\def\ref{\par\noindent\hangindent 20pt}
\def\mincir{\raise -2.truept\hbox{\rlap{\hbox{$\sim$}}\raise5.truept
\hbox{$<$}\ }}
\def\magcir{\raise -2.truept\hbox{\rlap{\hbox{$\sim$}}\raise5.truept
\hbox{$>$}\ }}
\def\gr{\kern 2pt\hbox{}^\circ{\kern -2pt K}} 
\def\asymp{\raise -4.3truept\hbox{$ \ \widetilde{\phantom{xy}} \ $}}

\def\hh{\hskip0.7truecm}
\def\h1{\hskip0.37truecm}
\def\s2{\hskip0.18truecm}
\def\w3{\hskip0.075truecm}

\null
\bigskip
\bigskip
\centerline{\cub Higher Order Moments of the Matter Distribution}
\medskip
\medskip
\centerline{\cub in Scale--Free Cosmological Simulations}
\medskip
\medskip
\centerline{\cub with Large Dynamic Range}
\bigskip
\bigskip
\bigskip
\bigskip
\centerline{{\bf Francesco Lucchin}$^1$, ~{\bf Sabino Matarrese}$^2$,}
\bigskip
\centerline{{\bf Adrian L. Melott}$^3$ and ~{\bf Lauro Moscardini}$^1$}
\bigskip
\bigskip
\bigskip
\medskip
\centerline{$^1$ {\it Dipartimento  di Astronomia, Universit\`a di Padova,}}

\centerline{{\it vicolo dell'Osservatorio 5, I--35122 Padova, Italy}}

\bigskip

\centerline{$^2$ {\it Dipartimento di Fisica ``G. Galilei", Universit\`a di
Padova,}}

\centerline{{\it via Marzolo 8, I--35131 Padova, Italy}}

\bigskip

\centerline{$^3$ {\it Department of Physics and Astronomy,}}

\centerline{{\it University of Kansas, Lawrence KS 66045}}

\vfill\eject

\vskip 1.truecm

\noindent{\bf Abstract.} We calculate reduced moments $\overline \xi_q$ of the
matter density fluctuations, up to order $q=5$, from counts in cells produced
by Particle--Mesh numerical simulations with scale--free Gaussian initial
conditions.  We use power--law spectra $P(k) \propto k^n$ with indices
$n=-3,~-2,~-1,~0,~1$. Due to the supposed absence of characteristic times or
scales in our models, all quantities are expected to depend on a single
scaling variable. For each model, the moments at all times can be
expressed in terms of the variance $\overline \xi_2$, alone.
We look for agreement with the hierarchical scaling ansatz, according to which
$\overline \xi_q \propto \overline \xi_2^{~q-1}$.
For $n\leq -2$ models we find strong deviations from the hierarchy, which
are mostly due to the presence of boundary problems in the
simulations. A small, residual signal of deviation from the hierarchical
scaling is however also found in $n \geq -1$ models.
For the first time, due
to our large dynamic range and careful checks of scaling and shot--noise
effects, we are able to detect evolution away from the perturbation theory
result.
\smallskip
\noindent{\it Subject headings:} Galaxies: formation, clustering --
large--scale structure of the Universe.

\vfill\eject

\noindent {\bf 1. Introduction}
\medskip

A fundamental problem in the analysis of the matter distribution in
the universe is to choose simple statistical tools able to provide the
most compact information on both the initial conditions and the subsequent
evolution of density fluctuations. In this sense, the study of count
probabilities has proved a useful statistical technique. They can be used to
follow the action of gravity
during the mildly nonlinear as well as fully nonlinear phases of structure
formation. The count probability approach, dating back to the early work
of Hubble (1934), has been recently applied to a number of galaxy samples:
Efstathiou et al. (1990)
calculated the variance of IRAS--selected galaxies in the QDOT sample
for roughly cubical cells of various sizes, while Loveday
et al. (1992) performed the same analysis in the Stromlo--APM
redshift survey; Saunders et al. (1991)
computed the skewness of density fluctuations, after smoothing the
QDOT galaxy distribution by a Gaussian filter.
A statistical analysis of the CfA
and SSRS optical galaxy samples in terms of moments of counts in cells
has been recently performed by Gazta\~naga (1992; see also
Gazta\~naga \& Yokoyama 1993). A more recent
analysis of this type, up to the fifth connected moment, has been performed
by Bouchet et al. (1993) on the $1.2$ Jy IRAS Galaxy Redshift Survey
(see also Bouchet, Davis, \& Strauss 1992).
Compared to connected correlation functions of order $q$,
$\xi_q({\bf x}_1, \dots, {\bf x}_q)$, reduced
moments (or {\it cumulants}) of the same order,
$\overline \xi_q$, of the fractional density
fluctuation enhance the signal--to--noise ratio, though at the expense of
reducing the amount of geometrical information.
One has the following connection between the above quantities
$$
\overline \xi_q(R) \equiv \int d^3 x_1 \dots d^3 x_q W_R({\bf x}_1)
\dots W_R({\bf x}_q) \xi_q({\bf x}_1, \dots, {\bf x}_q),
\eqno(1)
$$
where $W_R({\bf x})$ defines a suitable filter over a volume of size $R$.
These relations allow one to connect the results on moments
of galaxy counts in cells with the large amount of available data on galaxy
correlation functions.
Actually, moments of counts in cells can be related to the
$\xi_q$ only after shot--noise subtraction (see the following Section).
Early observations of higher order (i.e. $q>2$) correlation
functions established the validity of the so--called hierarchical scaling
ansatz according to which correlations of order $q$ can be expressed as
suitable sums of products of $q-1$ two--point functions (Groth \&
Peebles 1977; Fry \& Peebles 1978; Sharp, Bonometto, \& Lucchin 1984).
If the two--point function scales with distance as a power--law,
$\xi(r) \propto r^{-\gamma}$, or if the filtering radius is larger than the
typical correlation length, a related hierarchy holds for
the reduced moments,
$$
\overline \xi_q(R) = S_q \overline \xi_2^{~q-1}(R), \ \ \ \ \ \ \ \ q>2,
\eqno(2)
$$
with constant coefficients $S_q$. The hierarchical scaling of Eq.(2)
can be given a theoretical justification in two different regimes.
Starting from Gaussian density fluctuations, perturbation
theory shows that the action of gravity, already in the mildly
nonlinear regime, predicts the above hierarchical structure
(Peebles 1980; Fry 1984b; Goroff et al. 1986; Bernardeau 1992).
The hierarchical scaling also represents a self--consistent solution of the
BBGKY equations in the fully relaxed, highly nonlinear regime
(Davis \& Peebles 1977; Fry 1984a; Hamilton 1988).
The validity of this ansatz has been successfully tested
in numerical simulations by a number of authors: Coles \& Frenk (1991);
Bouchet \& Hernquist (1992, hereafter BH); Weinberg \& Cole (1992,
hereafter WC); Lahav et al. (1993, hereafter LIIS);
Fry, Melott, \& Shandarin (1993). Most of these works, however, deal with
the skewness, $\overline \xi_3$, vs. the variance, $\sigma^2 \equiv
\overline \xi_2$, relation (see also Silk \& Juszkiewicz 1991).
Indeed, much recent work has been devoted to a detailed analysis of the
skewness ratio $S_3 \equiv \overline \xi_3(R) / \overline \xi^{~2}_2(R)$. In
particular,
Juszkiewicz \& Bouchet (1992) and Juszkiewicz, Bouchet, \& Colombi (1993)
used a second order perturbation theory in Eulerian coordinates to
compute the dependence of $S_3$ on the type of window function as well
as on the spectral index $n$ for Gaussian density fluctuations with scale--free
spectra $P(k) \propto k^n$. Bouchet et al. (1992) used second order
perturbation expansion in Lagrangian space to evaluate the dependence
of $S_3$ on the density parameter $\Omega$.

The relevance of the primordial skewness of density fluctuations in determining
both the dynamical evolution and the present texture of the matter
distribution has been discussed by Moscardini et al. (1991), Messina
et al. (1992) and WC. The quantity $S_3$ has been calculated by Coles et al.
(1993) for the mass and galaxy distribution in $N$--body simulations of
skewed Cold Dark Matter (CDM) models and shown that this quantity
can be used as a powerful test to discriminate among various statistical
distributions of primordial fluctuations.

{}From the theoretical point of view, even if we accept that the mass
distribution follows the hierarchical law on scales affected by
nonlinear evolution, one still has to ask whether such a scaling is
stable against the nonlinear (and possibly non--local) biasing process
leading to the galaxy distribution. This issue has been partially
solved by Fry \& Gazta\~naga (1993), who showed that the hierarchical
scaling is indeed preserved by a rather general type of bias in the
limit of small fluctuations. Conversely, it might be
that the observed scaling of higher order correlation functions is
entirely due to the bias mechanism, instead of reflecting the true
statistical properties of the underlying matter distribution. Actually,
Politzer \& Wise (1984), for the Gaussian case, and Matarrese, Lucchin, \&
Bonometto (1986), for the non--Gaussian one, argued that
Eq.(2) is also implied on intermediate and large scales, if the
biasing mechanism requires a high--density threshold, while further terms
deriving from the Kirkwood expansion (e.g. Peebles 1980) also appear on
small scales. On the other hand, biasing with moderate threshold may
lead to the hierarchical form (Melott \& Fry 1986).
Any possible contamination of the hierarchical scaling law due to
redshift--space distortions has been
shown to be negligible by a number of authors (Bouchet et al. 1992; LIIS;
Coles et al. 1993; Juszkiewicz, Bouchet, \& Colombi 1993).

In this paper we test the validity of the hierarchical scaling law
of Eq.(2) for the third
(skewness), fourth (kurtosis) and fifth connected moments of the density
contrast in $N$--body simulations of the gravitational evolution of
scale--free Gaussian models with spectral index $n=-3,-2,-1,~0,~1$.
We also study the dependence of $S_q$ (for $q=3,~4,~5$) on the
primordial spectral index $n$.
Moments of counts in cells have already been computed in numerical simulations
for some of these models. Bouchet \& Hernquist (1992), besides
considering CDM and Hot Dark Matter models, run a tree code
with white--noise (i.e. $n=0$) initial conditions. In the frame of a
comparison of Particle--Mesh simulations with both Gaussian and
non--Gaussian initial conditions, WC calculate $S_3$ for
$n=-2,-1,~0$ initial spectra. Finally, LIIS
analyze counts in cells in tree code simulations for various
models, including $n=-1,~0,~1$ scale--free models.
Our analysis, besides considering a wider ensemble of power--law models,
covers a much larger dynamic range both in time and resolution. In Table I we
show the range in the {\it rms} fluctuation $\sigma$ and spectra studied for
our simulation
and those cited in this paragraph. It appears that ours is the only
study to date with sufficient dynamic range and control of boundary conditions
to reliably detect evolution away from the perturbation theory result. We also
have looked at a variety of pure power--law models, so that such dependence
can be detected.

The plan of the paper is as follows. In Section 2 we give the theoretical
background. Section 3 presents the numerical simulations and the
cautions used in the following analysis, while Section 4 discusses our
results on the analysis of the moments of counts
in cells. Conclusions are summarized in Section 5.

\bigskip
\noindent {\bf 2. Moments of Counts in Cells}
\medskip

The quantities $\overline \xi_q(R)$ defined above represent the reduced
moments of the density fluctuation
$\delta_R({\bf x}) \equiv \varrho_R({\bf x})/\overline \varrho -1$
(here $\varrho_R$ is the density smoothed over the scale $R$ and
$\overline \varrho$ its average); these can be
derived from the {\it moment generating function} ${\cal M}(s) \equiv
\int d \varrho_R {\cal P}(\varrho_R)
\exp(is \varrho_R/\overline \varrho)$, where ${\cal P}(\varrho_R)$
gives the probability density of the continuous variable $\varrho_R({\bf x})$.
One has
$$
\ln {\cal M}(s) = is + \sum_{q=2}^\infty{ (is)^q \over q!} \overline \xi_q(R).
\eqno(3)
$$
Given the probability density or the moment generating function,
one can easily generate discrete count--probabilities ${\cal P}_m$ (e.g.
Peebles 1980).
In fact, the count ${\cal P}_m$ can be understood as
the probability that, in a realization of the stochastic process
$\varrho_R$, $m$ objects are found in a randomly placed cell of
volume $V=Nv$, where $v$ is the specific volume $1/\overline \varrho$ and $N$
is the expected number of objects in that volume.
Consider then $\varrho_R$ as giving the mean density of an ensemble of local
Poisson count distributions with mean $N v \varrho_R$, whose
moment generating function is $\exp [Nv \varrho_R (e^{is} - 1)]$.
Note that the Poisson model does not necessarily provide a good
representation of discreteness effects (see e.g. Coles \& Frenk 1991;
Borgani et al. 1993); in particular, it
is likely to fail when the expected number of objects $N$ in the cell volume
is smaller than unity, i.e., when $V \ll v$.
Averaging over the $\varrho_R$ ensemble produces
the moment generating function for the discrete process as
$$
{\cal M}_{dis}(s) = \int d \varrho_R \ {\cal P}(\varrho_R) \ \exp\bigl[Nv
\varrho_R (e^{is} - 1)\bigr] = {\cal M}[-i(e^{is}-1)N]:
\eqno(4)
$$
the moment generating function of discrete counts, ${\cal M}_{dis}(s)$, is
obtained from the continuous one, ${\cal M}(s)$, by the replacement
$is \to (e^{is} -1)N$.
By inverse Fourier transforming ${\cal M}_{dis}(s)$ one gets
${\cal P}_{dis}(\varrho_R) = \sum_{m=0}^\infty \delta(\varrho_R - m \overline
\varrho) {\cal P}_m$, where $\delta$ is the Dirac delta--function and
the counts ${\cal P}_m$ are defined through a {\it Poisson transform} of
${\cal P}(\varrho_R)$,
$$
{\cal P}_m = \int d\varrho_R \ {\cal P}(\varrho_R) \ {(Nv\varrho_R)^m \over
m!} e^{-Nv\varrho_R}.
\eqno(5)
$$

One can then define the normalized central moments of the counts
in cells of volume $V$ as $\mu_q \equiv \langle
((m - \overline m)/\overline m)^q \rangle$, where
$$
\langle m^q \rangle = \sum_{m=1}^\infty m^q {\cal P}_m = (-i)^q
{d^q {\cal M}_{dis}(s) \over ds^q} \bigg\vert_{s=0},
\eqno(6)
$$
and $\overline m \equiv \sum_{m=1}^\infty m {\cal P}_m =N$.
For small $N$, at fixed moments of
$\varrho_R/\overline \varrho$, the count distribution is
shot--noise dominated and
${\cal M}_{dis}(s)$ reduces to the moment generating function of a Poisson
process with mean $N$: if the cell volume $V$ is too small,
statistical fluctuations dominate the realization and
one is unable to get any faithful statistical information on
${\cal P}(\varrho_R)$ from the counts.
This can be seen by writing ${\cal P}(\varrho_R)$ in Eq.(4) through its
Fourier transform and expanding in reduced moments $\overline \xi_q(R)$,
$$
{\cal M}_{dis}(s) = \int d y \ \exp\bigl[y(e^{is} -1)\bigr]
\int_{-\infty}^\infty
{d t \over 2\pi}
e^{i t(N-y)} \exp \biggl[\sum_{q=2}^\infty {(itN)^q \over q!}
\overline \xi_q(R) \biggr],
\eqno(7)
$$
which for small $N$ and fixed $\overline \xi_q(R)$ yields
$$
{\cal M}_{dis}(s) \approx \int d y \ \exp\bigl[y(e^{is} - 1)\bigr] \
\delta(y-N) =
\exp[N(e^{is} - 1)].
\eqno(8)
$$
When dealing with the $N \to 0$ limit above, one should, however, consider
the volume dependence of the connected moments. Actually, if the
hierarchical scaling of Eq.(2) holds and $\sigma^2 \propto
N^{-\gamma/3}$ as $N \to 0$, shot--noise dominates for small filtering scales,
provided that the effective spectral index $n_{eff} \equiv \gamma -3$
is smaller than zero.
Note that on small scales $n_{eff} \neq n$.

Conversely, shot--noise may even dominate for large cell sizes,
$N \to \infty$, where $\gamma \to n+3$, if the spectral index $n$
is larger than zero. Except for these cases, one generally expects
that the discrete counts $N {\cal P}_m$ reduce to the original continuous
distribution of $\varrho_R$, for large $N$ and $m$ and fixed $m/N$, as a
property of the Poisson transform. In fact, using the asymptotic
representation of the Poisson counts,
$$
{(Nv\varrho_R)^m \over m!} e^{-Nv\varrho_R} \sim
{1 \over \sqrt{2\pi N v\varrho_R}} \exp\biggl[ -
{(m-N v\varrho_R)^2\over 2N v\varrho_R} \biggr],
\eqno(9)
$$
in the integrand of Eq.(5), and taking
the limit for $N \to \infty$ one gets $N {\cal P}_m \sim
{\cal P}(\varrho_R)$, with $\varrho_R = \overline \varrho m/N$.

We can conclude that the optimal range of cell sizes $R\equiv V^{1/3}$
depends on the spectral slope $n$: in a numerical simulation such as ours, with
mean interparticle distance $\ell$, one should
require $R> {\rm Max}[\ell , ~\ell ~\sigma^{-2/3}(\ell)]$, for every $n$; for
$n>0$, however, one should also require $R < \ell ~\sigma^{-2/3}(R)$.

{}From Eqs.(3), (4) and (6) above one can explicitly find the required
relations among the moments of the counts and the reduced moments
of the continuous variable $\delta_R$. Inverting these relations, one finally
obtains the cumulants $\overline \xi_q$ as a function of the moments
of counts in cells $\mu_n$, up to order $q$.
We have, in particular,
$$
\eqalignno{
\overline \xi_2 & = \mu_2 - {1 \over N},
& (10) \cr
\overline \xi_3 & = \mu_3 - 3{\mu_2 \over N} + {2 \over N^2},
& (11) \cr
\overline \xi_4 & = \mu_4 - 6{\mu_3 \over N} - 3 \mu_2^2 +11{\mu_2 \over
N^2} - {6 \over N^3},
& (12) \cr
\overline \xi_5 & = \mu_5 - 10{\mu_4 \over N} - (10 \mu_2 - {35 \over N^2})
 \mu_3 + 30{\mu_2^2 \over N} - 50 {\mu_2 \over N^3} + {24 \over N^4}.
& (13) \cr}
$$

In what follows we shall also consider the ratios
$S_3 \equiv \overline \xi_3 /\overline \xi_2^{~2}$,
$S_4 \equiv \overline \xi_4 /\overline \xi_2^{~3}$ and
$S_5 \equiv \overline \xi_5 /\overline \xi_2^{~4}$, in order
to test whether the hierarchical scaling relations apply, i.e.,
whether these ratios are scale--independent, i.e. independent of the variance.

Theoretical predictions for the value of $S_3$ have been obtained using
second order perturbation theory. In order to get a consistent prediction for
$S_4$ and $S_5$ one respectively needs third and fourth order
perturbative calculations. Goroff et al. (1986) have computed these ratios
for initially Gaussian perturbations in standard CDM, by a clever
summation of tree diagrams. Filtering the density field by a Gaussian window
they obtain the values $S_3 \approx 3$, $S_4 \approx 16$ and $S_5 \sim 100$
on large scales, where the spectral slope tends to the Zel'dovich value $n=1$.
For scale--free Gaussian initial perturbations in an Einstein--de Sitter
model, Juszkiewicz, Bouchet, \& Colombi (1993) find the relation
$$
S_3^{(p)} = {34 \over 7} - (n+3), \ \ \ \ \ \ \ \ \ \ \ \ -3 \leq n < 1,
\eqno(14)
$$
using a spherical top--hat filter, while
for $n=1$ the perturbative prediction formally diverges. In the
latter case, taking into account that numerical simulations
cannot reproduce the initial spectrum above the Nyquist
frequency, they find $S_3^{(p)}(n=1) = 1.9$.
We shall compare these perturbative estimates
with our numerical results. The use of a sharp cubic filter instead of a
spherical one is not expected to introduce big changes in the $S_3$ vs. $n$
relation. Actually, we have numerically verified that the two filters
give essentially equivalent results, provided all quantities are compared at
equal smoothing volumes (see also, LIIS).

Perturbation theory implies an expansion of a series.
As is well known in basic physics, the series contains higher and higher
order powers of the perturbed quantity. It can only be expected to converge to
the correct result if this quantity is small. Indeed, perturbation theory is
going to fail as $\sigma$ gets of order unity, simply because the
gravitational field becomes arbitrarily large around regions of orbit mixing.
Thus perturbation theory ought to give better results for small $\sigma$,
and we can use that to compare with our procedures.
The value of numerical simulations such as these
is that we can investigate the nonlinear regime.

We believe that, for the first time, due to our large dynamic range and
careful checks of scaling, we are able to detect the evolution away from the
perturbation theory result.

\bigskip
\noindent {\bf 3. Numerical Simulations}
\medskip
The simulations studied herein are numerical models of the clustering of
collisionless matter in an expanding universe. We wish to investigate
the above scale--invariant behavior in the case of Gaussian initial conditions.
In order
to implement this, we use an $\Omega=1$ universe, as to choose otherwise would
introduce a preferred scale or time. We use pure power--law initial
perturbation spectra $P(k)\propto k^n$ with $n=-3,-2,-1,~0,~1$.
\medskip
The simulations are done with a Particle--Mesh (PM) code
(Hockney \& Eastwood 1981)
with $128^3$ particles in as many cells. In this paper we use 10 runs (two of
each spectral index) out of an ensemble of 50 generated for other systematic
studies (Melott \& Shandarin 1993). The PM code used in these studies is
highly optimized, using a staggered mesh scheme, and has twice the dynamical
resolution of any other PM code with which it has been compared [Melott 1986;
Melott, Weinberg, \& Gott 1988 (hereafter MWG); Weinberg 1993a,b]. Thus the
studies shown here are roughly equivalent in dynamic range to usual PM runs
with $128^3$ particles on a $256^3$ mesh except that we have less shot--noise
and collisionality. Having a relatively large number of particles has the
advantage of good mass resolution and the ability to impress initial
perturbations right up to the particle Nyquist frequency $k_{Ny}=64$. Other
methods such as P$^3$M and tree codes have not yet been able to run with
$128^3$ particles, which is relatively easy with PM, as can be seen by our
large ensemble of such runs. More details about the particular simulations
used here can be seen in Melott \& Shandarin (1993).
\medskip
Having stressed some advantages of our simulations, we would now like to
discuss
some of the precautions needed in using them, particularly for studies of
scale--free processes. Resolution is one problem.  Resolution has been stressed
as an advantage of codes in which short--range forces are calculated accurately
between point masses. In reality there are a number of different but related
kinds of resolution: (a) mass resolution, essentially the reciprocal of the
number of particles; (b) force resolution, essentially how accurately the force
law tracks $1/r^2$ at small separations; (c) spectral resolution, equivalent
to the minimum of the number of particles or Fourier analysis cells per unit
length, whichever is smaller;\break (d) minimum scale on which two--body
relaxation becomes important. PM methods are superior for all except (b), in
which other methods work better. We first consider this limitation.
\medskip
The growth rate of various modes in linear theory was studied in MWG for this
PM code. The growth rates for PM codes are usually described as being
unacceptable for $k>0.25 ~k_{Ny}$, or equivalently for
wavelength less than 8 cells. MWG found an equivalent performance for
$0.5~k_{Ny}$ to that found in usual PM codes at $0.25~k_{Ny}$. This results
from the staggered mesh scheme; it has since been confirmed in other
comparisons (Weinberg 1993a,b) and the PM code used here has been slightly
improved since that corresponding to a further 30\% resolution increase, {\it
i.e.} giving results at $\lambda=3$ cells equivalent to usual PM codes at
$\lambda=8$ cells.
\medskip
We still must take account of limited force resolution based on our grid
scheme. The advantage of the methods used here is that we can test for the
adequacy of our precautions by observing the results at various stages. In pure
power--law models, properties of the distribution should depend only on
$\sigma=(\delta\varrho/\varrho)_{rms}$, assuming the use of the identical
initial power--law
smoothing  windows. But $\sigma$ increases with time, and the
agreement on different scales at different times with the same
$\sigma$ is a strong consistency test. A similar strategy has already been used
   to
find previously unknown effects from the absence of waves larger than the
simulation volume (Kauffmann \& Melott 1992; Gramann 1992; Melott \&
Shandarin 1993).
\medskip
We first describe the stages of our simulations and then the restrictions we
applied. Our simulations were started by using the Zel'dovich (1970)
approximation, as first utilized by Klypin \& Shandarin (1983). It
is well--known
that this approximation is inaccurate beyond the time of nonlinearity [although
better than other approximations studied, with appropriate filtering; see
Coles, Melott, \& Shandarin (1993)] so the initial amplitude is well below
unity at the Nyquist frequency.
\medskip
The simulations were stopped at $k_{n\ell}=k_{Ny},\; 0.5~k_{Ny},\;
0.25~k_{Ny}, \dots, 2~k_f$ where $k_f=2\pi/L$ is the fundamental mode
of the box; $k_{n\ell}$ is defined by
$$
\langle\biggl({\delta \varrho \over \varrho} \biggr)^2\rangle_{k_{n\ell}}
\equiv \int\limits^{k_{n\ell}}_0 P(k) d^3k\equiv 1\; .
\eqno(15)
$$
In this study we have an available
range of $k_{n\ell}$ from $2~k_f$ to $64~k_f$. Everything in the simulations
should scale as $1/k_{n\ell}$, in the absence of boundary or resolution
problems. We make
the following restrictions on what scales will be studied. We will show results
based on counts in cells, in boxes of various sizes, at various stages of
nonlinearity. All results for a given spectral index will be plotted together
as a function of $\sigma=(\delta\varrho/\varrho)_{rms}$ and reveal
any problems.
\smallskip
\item{(a)} The stage with $k_{n\ell}=k_{Ny}$ will not be used to study
nonlinear effects since the
code is
known not to perform well at this frequency. We will begin with the stage
$k_{n\ell}=0.5~k_{Ny}=32~k_f$ in this study, except that we use the earlier
stage to help establish the linear limit.
\smallskip
\item{(b)} At each stage we will not present results for one pixel of density,
since this depends primarily on $k_{Ny}$. The smallest will be a cube of 8
such cells. We expect this will be acceptable since our code performs well at
wavelength of 4 cells, and collapse of $\lambda=4$ perturbations will
initially give rise to condensations of diameter 2. We will confirm this later.
\smallskip
\item{(c)} Kauffmann \& Melott (1992) found that for voids of size greater
than size $0.25 L$ self--similarity was broken in a model equivalent to
our index $n=-1$; see also Gramann (1992) and Melott \& Shandarin (1993). We
therefore restrict ourselves to cubes of
size $L/8$ or smaller. Combined with restriction (b), this leaves us with cells
for scale from $L/64$ to $L/8$. We do not expect this to work with
$n\leq-2$, and believe that for these values all so--called ``N--body"
simulations are at best crude. It would not be entirely a joke to say that in
this case a model could never be big enough to be a fair sample of itself.
Fortunately, it appears that the power spectrum of the universe turns over to
$n>0$ on large scales.
\medskip
\item{(d)} We use counts in cells, rather than the usual Cloud--in--Cell method
(Hockney \& Eastwood 1981) to bin densities. This procedure increases
shot--noise as compared to that present when the PM code calculates the
gravitational potential. For this reason
we do not use cell sizes with $\sigma<0.1$. In practice, in our simulations
this restriction eliminates cells where the shot--noise power
is comparable to the fluctuation power impressed on the simulation,
which makes the subtraction doubtful. Following our discussion in Section 2,
this only affects the largest cells at early times in our models with $n\geq
0$.
\medskip
Taken together, these restrictions can eliminate problems and give us much more
usable dynamic range than has been possible before in such a study. We verify
that it works remarkably well by examining the agreement between various stages
at the same $\sigma$.  It suggests that scale dependence found in previous
studies was probably a result of boundary problems. For $n\leq-2$ we find very
poor agreement, as expected.

As an illustration of the different appearance of the models, Figure 1 shows,
for the five spectra, grey--scale plots of $L/64$  thickness
slices at the stage corresponding to $k_{n\ell}=8$.

\bigskip
\noindent {\bf 4. Analysis and Results}
\medskip

In our scale--free simulations we expect that every quantity
depends on a single scaling variable, namely the variance $\overline \xi_2$.
Therefore we can plot the results of counts in cells at all times together
as a function of the variance. Figure 2 shows the skewness $\overline \xi_3$
vs. the variance, for all models, i.e. for all values of $n$.
The points shown are the average of the two runs; the corresponding dispersion
is always small and will not be shown here.
The dashed lines are the second order perturbative predictions [obtained from
Eq.(14)], for the same value of $n$; these computations derive from using a
spherical top--hat window, but we checked that changing from
spherical to cubic filter produces essentially the same results.
The solid line shows the two--parameter linear fit
$$
\log \overline \xi_q = A_q + B_q \log \overline \xi_2
\eqno(16)
$$
(for $q=3$); the corresponding coefficients and related
errors are reported in Table II. Figures 3 and 4, respectively, show the
kurtosis $\overline \xi_4$ and the fifth moment $\overline \xi_5$ vs.
$\overline \xi_2$, for all values of the initial spectral index.
Perturbative predictions are not available in these cases.
The solid lines represent the results
of linear fits from Eq.(16); best--fit coefficients and errors are reported
in Table III and IV, for the fourth and fifth moment, respectively.
Note that the $n\leq -2$ results are clearly not reliable, being highly
affected by the finite box--size (see Fry, Melott, \& Shandarin 1993).
In all other cases ($n=-1,~0,~1$), the scaling of
$\overline \xi_q$ ($q=3,~4,~5$) with
the variance is quite close to the hierarchical form,
$B_q = q-1$. Nevertheless, we detect a residual deviation
from this scaling, at more than three standard deviations for both the
skewness and the kurtosis, while for the fifth moment this result appears
at only one standard deviation. We consider rather unlikely that such a
deviation is produced by boundary effects.
Due to the large dynamic range of our simulations and to the careful
treatment of resolution and shot--noise problems, we are led to trust
these deviations, even though our results are closer to the hierarchy
than previous works (BH and LIIS), whose simulations are more affected
than ours by finite sample and resolution effects. For instance,
for initial white--noise, BH find $B_3=2.10 \pm 0.01$,
$B_4= 3.26 \pm 0.02$ and $B_5= 4.44 \pm 0.04$, while LIIS obtain
$B_3= 2.08 \pm 0.01$ and $B_4= 4.16 \pm 0.03$.
Note however, that, according to our previous discussion on shot--noise,
the highest $\sigma$ points appearing in BH, which come
from cells with size smaller than the mean interparticle distance,
are likely to be dominated by discreteness effects, which
decreased their statistical reliability.

In order to better display possible deviations from scale--invariance
we also plot the coefficients $S_q$ as a function of $\sigma$.
In Figures 5, 6 and 7 we show $S_3$, $S_4$ and $S_5$,
for all values of $n$. The meaning of the solid
lines and dashed ones (when present) is as before. Note that the trend of
$S_q$ with $n$ is the
same for all $q$: $S_q$ decreases with increasing $n$. Although this
qualitative trend is the same predicted by perturbation theory
for $q=3$,
the value of the coefficients is only in partial agreement with it.
In Tables II -- IV, we report the coefficients of
the one--parameter fits of $S_q$, obtained at fixed hierarchical slope;
dotted lines in Figures 5 -- 7 represent these best--fit coefficients.
Note that the values of $S_3$ are generally different from $S_3^{(p)}$ as
given by Eq.(14); however, since perturbation theory is at most consistent
with mildly nonlinear evolution, we also estimated the skewness ratio by
fitting only points with $\sigma \leq 0.7$: this is reported as $S_3^\star$
in Table II (we similarly define $S_4^\star$ and $S_5^\star$ in Tables
III and IV). The agreement with perturbation
theory is indeed improved, but we still get slightly higher values:
$S_3^\star = 4.5, ~3.4,~3.1,~2.0,~1.7$, instead of $S_3^{(p)}=
4.9, ~3.9,~2.9,~1.9,~1.9$, for $n =-3,~-2,~-1,~0,~1$, respectively.
In deriving the values of $S_3^\star$ above, we also used an earlier
stage of the simulations corresponding to $k_{n\ell} = 64 ~k_f$.
For their simulations with
$n=0$, BH argue that $S_3$ should tend to about $1.8$ at large scales, very
close to the perturbative prediction; their result is obtained from large
cell sizes, corresponding to about one quarter of the computational box size,
where sample effects make the resulting data less reliable.

Let us stress that agreement with second order perturbation theory is not
a good test of the quality of $N$--body data, except in the linear regime.
However, the behavior of such codes has already been widely tested in
this regime.
In the nonlinear regime, one should use $N$--body data to test the
reliability of perturbation expansion results, such as the second order
estimate of $S_3$ reported in Eq.(14). In this sense one can consider a
success that the qualitative trend of $S_3$ with $n$ is correctly predicted
by perturbation theory! It would be interesting
to have similar predictions for higher order moments, to compare with our
numerical results.

\bigskip
\noindent {\bf 5. Conclusions}
\medskip

We have studied the properties of higher order moments of the matter
distribution generated by gravity in the nonlinear regime. This was done by
analyzing numerical simulations of the evolution of initially Gaussian
perturbations with scale--free power--spectra (with spectral index
$n=-3,~-2~,-1,~0,~1$) at many evolution stages and smoothing scales. Our
results for $n\geq -1$ models indicate that these moments are fairly close to
the hierarchical scaling ansatz, according to which connected moments
$\overline \xi_q$, of order $q$, are proportional to the $q-1$ power of
the variance $\overline \xi_2$. However, we detect a residual
dependence, above the statistical noise, of the coefficients
$S_q = \overline \xi_q/\overline \xi_2^{~q-1}$ on the
variance, i.e. on scale. In order to detect such a signal from
the data we had to properly account for shot--noise, finite resolution effects
and boundary problems. For models with $n\leq - 2$, where the amount of
large--scale power does not allow a fair representation of low frequency
modes of the density field (due to the finite box size), a stronger
dependence of $S_q$ on scale is found.

The skewness ratio $S_3$ is found to decrease with increasing
spectral slope, $n$, i.e. with decreasing large--scale power, as correctly
predicted by perturbative calculations, although our values for $S_3$
are only consistent with the theory if the small $\sigma$ limit of these
quantities is considered. The same qualitative trend with $n$ is seen for
the higher order coefficients $S_4$ and $S_5$.

Altogether, these results indicate that: 1) the dynamical effect of gravity is
such as to generate non--Gaussian signatures on a Gaussian initial density
field, already in the earliest stages of evolution and/or on large
scales; 2) the hierarchical ansatz provides only an approximate description
of the behavior of
higher order moments of the density fluctuation field with the scale.

\bigskip
\bigskip
\noindent {\bf Acknowledgements}
\medskip
Computations were performed at the National Center for Supercomputing
Applications, Urbana, Illinois. ALM gratefully acknowledges the support of NSF
grants AST--9021414, NSF EPSCoR Grant number OSR--9255223, and NASA grant
NAGW--2923. FL, SM, and LM thank the Ministero Italiano  dell'Universit\`{a} e
della Ricerca Scientifica e Tecnologica for financial support. We would like
to thank Paolo Catelan and David Weinberg for fruitful discussions.
\vfill\eject

\bigskip
\centerline{\bf REFERENCES}
\medskip

\ref {Bernardeau, F. 1992, ApJ, 392, 1}

\ref {Borgani, S., Coles, P., Moscardini, L., \& Plionis, M. 1993,
MNRAS, submitted}

\ref {Bouchet, F.R., Davis, M., \& Strauss, M.A. 1992, in
Proc. of the 2nd DAEC Meeting on The Distribution of Matter
in the Universe, eds. G.A. Mamon, \& D. Gerbal, p. 287}

\ref {Bouchet, F.R., \& Hernquist, L. 1992, ApJ, 400, 25 (BH)}

\ref {Bouchet, F.R., Juszkiewicz, R., Colombi, S., \& Pellat, R. 1992,
ApJ, 394, L5}

\ref {Bouchet, F.R., Strauss, M.A., Davis, M., Fisher, K.B., Yahil, A.,
\& Huchra, J.P. 1993, ApJ, submitted}

\ref {Coles, P., \& Frenk, C.S. 1991, MNRAS, 253, 727}

\ref {Coles, P., Melott, A.L., \& Shandarin, S.F. 1993, MNRAS, 260, 75}

\ref {Coles, P., Moscardini, L., Lucchin, F., Matarrese, S., \& Messina, A.
1993, MNRAS, submitted}

\ref {Davis, M., \& Peebles, P.J.E. 1977, ApJS, 35, 425}

\ref {Efstathiou, G., Kaiser, N., Saunders, W., Lawrence, A.,
Rowan--Robinson, M., Ellis, R.S., \& Frenk, C.S. 1990, MNRAS, 247, 10P}

\ref {Fry, J.N. 1984a, ApJ, 277, L5}

\ref {Fry, J.N. 1984b, ApJ, 279, 499}

\ref {Fry, J.N., \& Gazta\~naga, E. 1993, ApJ, submitted}

\ref {Fry, J.N.,  Melott, A.L., \& Shandarin, S. 1993, ApJ, in press}

\ref {Fry, J.N., \& Peebles, P.J.E. 1978, ApJ, 221, 19}

\ref {Gazta\~naga, E. 1992, ApJ, 398, L17}

\ref {Gazta\~naga, E., \& Yokoyama, J. 1993, ApJ, 403, 450}

\ref {Goroff, M.H., Grinstein, B., Rey, S.--J., \& Wise, M. 1986, ApJ, 311, 6}

\ref {Gramann, M. 1992, ApJ, 401, 19}

\ref {Groth, E.J., \& Peebles, P.J.E. 1977, ApJ, 217, 385}

\ref {Hamilton, A.J.S. 1988, ApJ, 332, 67}

\ref {Hockney, R., \& Eastwood, J.W. 1981, Computer Simulation Using Particles
(New York: McGraw--Hill)}

\ref {Hubble, E.P. 1934, ApJ, 79, 8}

\ref {Juszkiewicz, R., \& Bouchet, F.R. 1992, in
Proc. of the 2nd DAEC Meeting on The Distribution of Matter
in the Universe, eds. G.A. Mamon, \& D. Gerbal, p. 301}

\ref {Juszkiewicz, R., Bouchet, F., \& Colombi, S. 1993, ApJ, in press}

\ref {Kauffmann, G., \& Melott, A.L. 1992, ApJ, 393, 415}

\ref {Klypin, A.A., \& Shandarin, S.F. 1983, MNRAS, 104, 891}

\ref {Lahav, O., Itoh, M., Inagaki, S., \& Suto, Y. 1993, ApJ, 402, 387
(LIIS)}

\ref {Loveday, J., Efstathiou, G., Peterson, B.A., \& Maddox, S.J. 1992,
ApJ, 400, L43}

\ref {Matarrese, S., Lucchin, F., \& Bonometto, S.A. 1986, ApJ, 310, L21}

\ref {Melott, A.L. 1986, Phys. Rev. Lett., 56, 1992}

\ref {Melott, A.L., \& Fry, J.N. 1986, ApJ, 305, 1}

\ref {Melott, A.L., \& Shandarin, S.F. 1993, ApJ, in press}

\ref {Melott, A.L., Weinberg, D.H., \& Gott, J.R. 1988, ApJ, 328, 50 (MWG)}

\ref {Messina, A., Lucchin, F., Matarrese, S., \& Moscardini, L. 1992,
Astroparticle Phys., 1, 99}

\ref {Moscardini, L., Matarrese, S., Lucchin, F., \& Messina, A.
1991, MNRAS, 248, 424}

\ref {Peebles, P.J.E. 1980, The Large Scale Structure of the Universe
(Princeton: Princeton Univ. Press)}

\ref {Politzer, H.D., \& Wise, M.B. 1984, ApJ, 285, L1}

\ref {Saunders, W., et al. 1991, Nature, 349, 32}

\ref {Sharp, N., Bonometto, S.A., \& Lucchin, F. 1984, A\&A, 130, 79}

\ref {Silk, J., \& Juszkiewicz, R. 1991, Nature, 353, 386}

\ref {Weinberg, D.H. 1993a, in preparation}

\ref {Weinberg, D.H. 1993b, personal communication}

\ref {Weinberg, D.H., \& Cole, S. 1992, MNRAS, 259, 652 (WC)}

\ref {Zel'dovich, Ya.B. 1970, A\&A, 5, 84}

\vfill\eject
\centerline{\bf Figure captions}
\bigskip
\bigskip
\noindent
{\bf Figure 1}. Grey--scale plots of the projected density in slices of
thickness $L/64$ at stage $k_{n\ell}=8 k_f$ in our simulations. Regions below
the mean density are white; regions of density $\rho>10$ are black; a grey
scale is used in between. (a) $n=-3$; (b) $n=-2$; (c) $n=-1$; (d) $n=0$;
(e) $n=1$.
\medskip
\noindent
{\bf Figure 2}. The skewness $\overline \xi_3$ vs. the variance
$\overline \xi_2$ for the different models. The solid lines represent the
best--fit obtained from Eq.(16), while
the dashed lines are the perturbative prediction.
Different symbols refer to different stages of the simulations:
$k_{n\ell}=32 ~k_f$, filled triangles; $k_{n\ell}=16 ~k_f$, asterisks;
$k_{n\ell}=8 ~k_f$, open triangles; $k_{n\ell}=4 ~k_f$, open squares;
$k_{n\ell}=2 ~k_f$, open circles.
Note that circles are absent for $n=-3$; they were left out due to lack of
any benefit in including them.
\medskip
\noindent
{\bf Figure 3}. The kurtosis $\overline \xi_4$ vs. the variance
$\overline \xi_2$ for the different models. The solid lines represent the
best--fit obtained from Eq.(16).
Different symbols refer to different stages of the simulations as in Figure 2.
\medskip
\noindent
{\bf Figure 4}. The fifth connected moment $\overline \xi_5$ vs. the variance
$\overline \xi_2$ for the different models. The solid lines represent the
best--fit obtained from Eq.(16).
Different symbols refer to different stages of the simulations as in Figure 2.
\medskip
\noindent
{\bf Figure 5}. The skewness coefficient $S_3 = \overline \xi_3/\overline
\xi_2^{~2}$ vs. the {\it rms} fluctuation
$\sigma$ for the different models. The solid lines represent the
best--fit obtained from Eq.(16), the dashed lines are the perturbative
prediction, finally the dotted lines are the result of a best--fit forced to
the hierarchical slope. Different symbols refer to different stages of the
simulations as in Figure 2.
\medskip
\noindent
{\bf Figure 6}. The kurtosis coefficient $S_4 = \overline \xi_4/\overline
\xi_2^{~3}$ vs. the {\it rms} fluctuation
$\sigma$ for the different models. The solid lines represent the
best--fit obtained from Eq.(16), while
the dotted lines are the result of a best--fit forced to
the hierarchical slope. Different symbols refer to different stages of the
simulations as in Figure 2.
\medskip
\noindent
{\bf Figure 7}. The fifth moment coefficient $S_5 = \overline \xi_5/\overline
\xi_2^{~4}$ vs. the {\it rms} fluctuation $\sigma$ for the different models.
The solid lines represent the best--fit obtained from Eq.(16), while the
dotted lines are the result of a best--fit forced to
the hierarchical slope. Different symbols refer to different stages of the
simulations as in Figure 2.
\vfill\eject

\centerline{\bf Table I}
\smallskip
\centerline{\bf Comparison of Dynamic Range in Recent Similar Studies}
\centerline{\bf Ratio of $\sigma_{max}$ to $\sigma_{min}$}
\bigskip
\bigskip
{\settabs 5 \columns
\+  & ~~BH$^*$ & WC$^{**}$ & LIIS$^{***}$ & this study  \cr
\+ $n=-3$ &  &      & & ~~$\sim 10$ \cr
\+ $n=-2$ &  & $1.86$ & & ~~$\sim 20$ \cr
\+ $n=-1$ &  & $1.86$ & $\sim 8$ & ~~$\sim 64$ \cr
\+ $n=$\h1 $0$ & $\sim 100$ & $1.86$ & $\sim 15$ & ~~$\sim 100$ \cr
\+ $n=+1$ & & & $\sim 30$ & ~~$\sim 100$ \cr}
\bigskip
\item{*} estimated from BH Figure 8.
\item{**} estimated from WC Table 2. WC did not attempt to study the evolution
of these moments over a wide dynamic range.
\item{***} numbers estimated from LIIS Figure 2. Discrepancies up to a
factor two exist between results at different moments (see LIIS Table I).

\bigskip
\bigskip
\bigskip
\bigskip
\baselineskip=18truept
\centerline{\bf Table II}
\smallskip
\centerline{\bf Third moment coefficients}
\bigskip
{\settabs 6 \columns
\+ &\hh $A_3$ &\hh $B_3$ & $S_3$ & $S_3^\star$ & $S_3^{(p)}$ \cr
\+         &       &       &         &       &      \cr
\+ $n=-3$ & $0.74\pm 0.01$ & $2.28\pm 0.02$ & $7.0$ & $4.5$ & $4.9$ \cr
\+ $n=-2$ & $0.57\pm 0.01$ & $2.07\pm 0.01$ & $4.0$ & $3.4$ & $3.9$ \cr
\+ $n=-1$ & $0.51\pm 0.01$ & $2.03\pm 0.01$ & $3.3$ & $3.1$ & $2.9$ \cr
\+ $n=$\h1 $0$ & $0.38\pm 0.01$ & $2.04\pm 0.01$ & $2.5$ & $2.0$ & $1.9$ \cr
\+ $n=+1$ & $0.30\pm 0.01$ & $2.04\pm 0.01$ & $2.0$ & $1.7$ & $1.9^\#$ \cr}

\bigskip\noindent
$\#$ see discussion in the text.

\vfill\eject
\centerline{\bf Table III}
\smallskip
\centerline{\bf Fourth moment coefficients}
\bigskip
{\settabs 5 \columns
\+  & \hh $A_4$ & \hh $B_4$ & \w3 $S_4$ & \w3 $S_4^\star$  \cr
\+     &       &         &        &     \cr
\+ $n=-3$ & $1.66\pm 0.02$ & $3.59\pm 0.05$ & $97.4$ & $29.5$ \cr
\+ $n=-2$ & $1.33\pm 0.02$ & $3.15\pm 0.03$ & $28.1$ & $19.1$ \cr
\+ $n=-1$ & $1.19\pm 0.02$ & $3.07\pm 0.02$ & $18.1$ & $16.0$ \cr
\+ $n=$\h1 $0$ & $0.96\pm 0.02$ & $3.04\pm 0.01$ & $10.0$ & $~9.4$ \cr
\+ $n=+1$ & $0.73\pm 0.02$ & $3.08\pm 0.02$ & $~6.2$ & $~5.9$ \cr}

\bigskip
\bigskip
\bigskip
\bigskip
\bigskip
\bigskip
\bigskip
\centerline{\bf Table IV}
\smallskip
\centerline{\bf Fifth moment coefficients}
\bigskip
{\settabs 5 \columns
\+  & \hh $A_5$ & \hh $B_5$ & \s2 $S_5$ & \w3 $S_5^\star$ \cr
\+     &      &         &        &     \cr
\+ $n=-3$ & $2.65\pm 0.04$ & $4.92\pm 0.08$ & $2091$ & $218$ \cr
\+ $n=-2$ & $2.15\pm 0.04$ & $4.24\pm 0.05$ & $~319$ & $161$ \cr
\+ $n=-1$ & $2.00\pm 0.03$ & $4.05\pm 0.03$ & $~147$ & $135$ \cr
\+ $n=$\h1 $0$ & $1.64\pm 0.03$ & $4.02\pm 0.03$ & $~~63$ & $~86$ \cr
\+ $n=+1$ & $1.31\pm 0.03$ & $4.02\pm 0.03$ & $~~30$ & $~48$ \cr}

\vfill\eject
\bye